\documentclass[]{emulateapj}
\usepackage{amsmath}
\usepackage{times}
\usepackage{graphics, subfigure}
\usepackage{epsfig}
\usepackage{multirow}
\usepackage{ulem}
\usepackage{pdfpages}

\def\simlt{\mathrel{\hbox{\rlap{\hbox{\lower4pt\hbox{$\sim$}}}\hbox{$<$}}}}
\def\simgt{\mathrel{\hbox{\rlap{\hbox{\lower4pt\hbox{$\sim$}}}\hbox{$>$}}}}
\def\Rsh{R_{\rm sh}}
\newcommand{\Vsh}{\dot{R}_{\rm sh}}
\newcommand{\zsh}{z_{\rm sh}}
\newcommand{\Vzsh}{\dot{z}_{\rm sh}}

\begin{document}

\title{EARLY HYDRODYNAMIC EVOLUTION OF A STELLAR COLLISION}

\author{Doron Kushnir\altaffilmark{1} and Boaz Katz\altaffilmark{1,2}} \altaffiltext{1}{Institute for Advanced Study, Einstein Drive, Princeton, New Jersey, 08540, USA} \altaffiltext{2}{John N.\ Bahcall Fellow}

\begin{abstract}
The early phase of the hydrodynamic evolution following collision of two stars is analyzed. Two strong shocks propagate at a constant velocity (which is a small fraction of the velocity of the approaching stars) from the contact surface toward the center of each star. The shocked region near the contact surface has a planar symmetry and a uniform pressure. The density vanishes at the (Lagrangian) surface of contact and the speed of sound diverges there. The temperature, however, reaches a finite value, since as the density vanishes, the finite pressure is radiation dominated. For Carbon-Oxygen white dwarfs collisions this temperature is too low for any appreciable nuclear burning at early times. The divergence of the speed of sound limits numerical studies of stellar collisions, as it makes convergence tests exceedingly expensive unless dedicated schemes are used. We provide a new one-dimensional Lagrangian numerical scheme to achieve this. Self-similar planar solutions are derived for zero-impact parameter collisions between two identical stars, under some simplifying assumptions. These solutions provide rough approximations that capture the main features of the flow and allow a general study as well as a detailed numerical verification test problem. The self-similar solution in the upstream frame is the planar version of previous piston problems that were studied in cylindrical and spherical symmetries. We found it timely to present a global picture of self similar piston problems. In particular, we derive new results regarding the non trivial transition to accelerating shocks at sufficiently declining densities (not relevant for collisions).
\end{abstract}
% -------------------------- End of abstract -----------------------

% -----------------------------------------------------------------------
% --------------------------  Sec 1: INTRODUCTION -----------------------
% -----------------------------------------------------------------------

\section{Introduction}
\label{sec:Introduction}

It was recently argued \citep{katz2012rate} that as many as $\sim1\%$ of all stars may collide with each other during their lifetime, due to the dynamics of typical field triple systems. Especially interesting is that the rate of collision between white dwarfs (WDs) in such systems may be as high as the rate of type Ia supernovae (SNe Ia). Although collisions of WDs were earlier believed to have rates which are orders of magnitude smaller than the rate of SNe Ia, they motivated three-dimensional hydrodynamic simulations of such collisions and the possible resulting thermonuclear explosion \citep{Benz1989,Raskin2009oti,Rosswog2009cwd,Loren-Aguilar2010,Raskin2010pdd,Hawley2012zip}. While the amount of $^{56}$Ni \citep[the decay of which powers the observed light,][]{Colgate69} synthesized in most of these simulations was non-negligible, the results were contradictory, with inconsistent amounts of $^{56}$Ni and different ignition sites of a detonation wave for the same initial conditions. This discrepancy was resolved by \citet{Kushnir2013}, where high resolution two-dimensional (2D) simulations with a fully resolved ignition process were employed \footnote{We note that the recent simulations preformed by \citet{Garcia2013} are at significant lower resolutions than those performed by \citet{Kushnir2013}.}. Moreover, it was shown that there is a strong correlation between the $^{56}$Ni yield and the total mass of colliding Carbon-Oxygen (CO) WDs (insensitive to their mass ratio) which spans the observed range of SNe Ia yields for the observed range of CO WDs masses. In all collisions the nuclear detonation is due to a well understood shock ignition, devoid of the commonly introduced free parameters such as the deflagration velocity or transition to detonation criteria. The detonation triggered by the collisions results in explosions which match key observational properties of SNe Ia. We believe that this is the main channel for these explosions.

In this paper we analyze the early hydrodynamic evolution of zero-impact parameter collisions between two identical stars. Our results are applicable to a wide variety of stellar collisions, and in particular they clarify the early evolution in the case of collisions between CO WDs. We restrict our analysis to early times in which the velocity of the approaching stars, $\pm v_{0}$, is roughly constant. The velocity is increasing since the stars accelerate towards each other in the gravitational field of each star, $g_{0}\simeq GM_{\star}/R_{\star}^{2}$, where $M_{\star}$ is the stellar mass and $R_{\star}$ is the stellar radius. In what follows, we limit our analysis to early times, for which $t\ll t_{0}\equiv v_{0}/g_{0}$.

The approaching velocity $v_{0}$ is much larger than the speed of sound near the stellar edge and immediately after contact two strong shock waves form that propagate from the contact surface towards the center of each of the stars. As we show, the shock velocity in the collision frame, $\Vzsh$, is much smaller than $v_{0}$ for stellar collisions with  $\Vzsh\lesssim v_{0}/7$. In the context of collisions between CO WDs this property leads to a lack of any appreciable nuclear burning at early times, $t\lesssim 0.1 v_{0}/g_{0}$. At later times, when a detonation is ignited \citep[typically at $t\sim t_{0}$, see e.g.][]{Kushnir2013}, there is already a significant amount of shocked material, which allows an efficient synthesis of $^{56}$Ni, consistent with SNe Ia observations. A detailed study of these detonations at later times is beyond the scope of this paper and will be described in a subsequent publication \citep{Kushnir2013b}.

The paper is organized as follows. In Section~\ref{sec:ContactProp} we discuss the properties of the contact region in an example of collision between CO WDs. We show that the contact region has a planar symmetry, a uniform pressure, a diverging speed of sound, and a finite temperature. These main features do not depend on the assumptions that the stars have equal masses, that the impact parameter is zero or on the specific density profile. The planar symmetry allows us to verify that the results of a realistic 2D simulation of a collision (with a limited resolution near the contact surface) are correct, by comparing it to a high resolution one-dimensional (1D) simulation. A numerical difficulty arises due to the diverging speed of sound and is overcome using a new 1D numerical scheme. In Section~\ref{sec:SES}, we consider the ideal case of zero impact collisions of identical stars with a (pre-collision) power-law density profile, $\rho=Kr^{-\omega}$, and an ideal equation of state (the somewhat confusing notations, $r$ for the distance from the contact surface and $\omega<0$ for increasing profile, are used to be consistent with previous literature on propagating shocks in cylindrical and spherical symmetries, see below). The exact self-similar solution of this problem as well as a simple and accurate analytical approximation are derived. We find excellent agreement between the 1D numerical results and the self-similar solutions, which shows that the self-similar solutions are achieved for this flow and that our numerical scheme solves the flow equations accurately.

The self-similar solution in the upstream frame is the planar version of previous piston problems that were studied in cylindrical and spherical symmetries \citep[][and references therein]{Sedov45,Taylor46,SedovBook}. We found it timely to present a global picture of self similar piston problems. In Section~\ref{sec:self_similar_Appendix}, the solutions for planar, cylindrical and spherical symmetries, for all values of the density power law index $\omega$ are presented and compared. In particular, we derive new results regarding the non trivial transition to accelerating shocks at sufficiently declining densities (not relevant for collisions) and point out interesting similarities and differences with the strong explosion problem.
% ---------------------------- End  of sec 1 ----------------------------

% -----------------------------------------------------------------------
% --------------------- Sec 2: Example + general properties ---------------
% -----------------------------------------------------------------------

\section{The contact region has a planar symmetry, a uniform pressure, a diverging speed of sound, and a finite temperature}\label{sec:ContactProp}

In this section we discuss the properties of the contact region. We begin with an example of collision between CO WDs (Section~\ref{sec:example}). We show that the contact region has a planar symmetry, a uniform pressure, a diverging speed of sound, and a finite temperature. These main features are general (Section~\ref{sec:general}), and do not depend on the assumptions that the stars have equal masses, that the impact parameter is zero or on the specific density profile. The planar symmetry allows us to verify that the results of a realistic 2D simulation of a collision (with a limited resolution near the contact surface) are correct, by comparing it to a high resolution 1D simulation (Section~\ref{sec:1D}). A numerical difficulty arises due to the diverging speed of sound and is overcome using a new 1D numerical scheme. Finally, we show in Section~\ref{sec:NoBurn} that for all CO WD collisions the temperature at the contact surface is generally too small for any appreciable nuclear burning at early times $t\lesssim 0.1 v_{0}/g_{0}$.

\subsection{Example: early stages of the collision of two CO WDs}
\label{sec:example}

\begin{figure}
\epsscale{1} \plotone{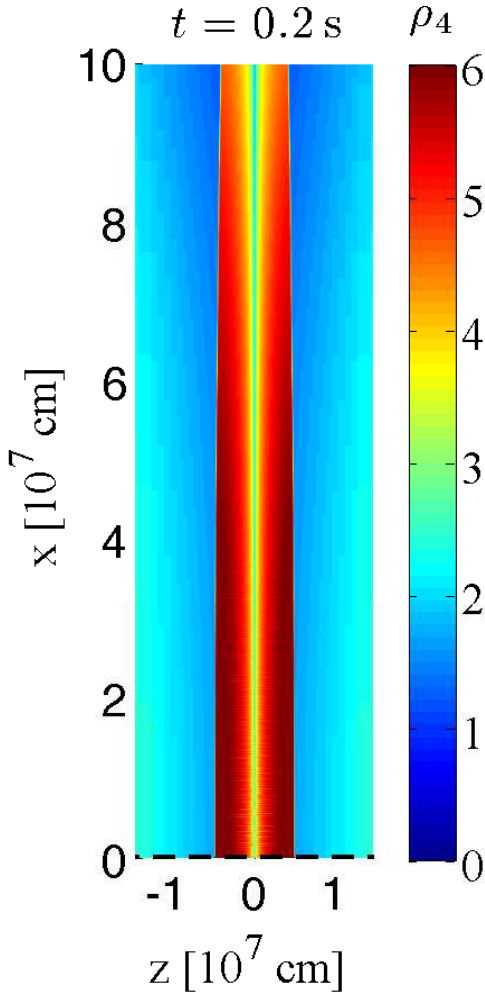} \caption{A density map at $t=0.2\,\textrm{s}$ from a 2D FLASH4.0 simulation with $\simeq1\,\textrm{km}$ resolution of the zero-impact-parameter collision of 0.64-0.64 $M_{\odot}$ CO WDs, previously moving on a Keplerian orbit ($\rho_{4}\equiv\rho/10^{4}\,\textrm{g}\,\textrm{cm}^{-3}$). The contact surface is at $z=0$. The positions of the shocks on the symmetry axis ($x=0$) are $\zsh\approx \pm0.5\times10^7\,\textrm{cm}$. The profiles of the flow variables on the symmetry axis (dashed lines) are given in Figure~\ref{fig:FlashProf}.}
\label{fig:Flash}
\end{figure}

\begin{figure}
\epsscale{1} \plotone{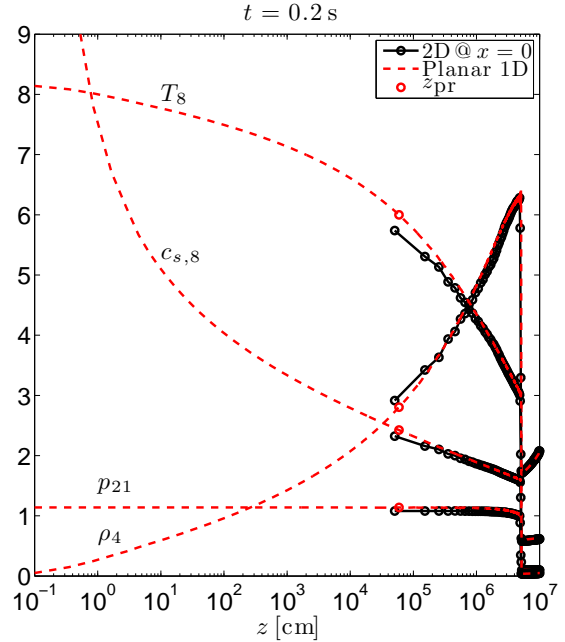} \caption{The profiles of the flow variables (temperature, speed of sound, pressure, and density) as function of the distance from the contact surface, $z$ from the 2D FLASH4.0 simulation of colliding CO WDs shown in Figure~\ref{fig:Flash} (black, the profiles are shown for $x=0$, which correspond to the dashed lines in Figure~\ref{fig:Flash}). The hydrodynamic profiles from a 1D planner numerical scheme (red, see text for details) are shown to agree (to better than $5\%$) with the 2D profiles. The outer position of the uniform pressure region at that time, $z_{\textrm{pr}}$, is marked with red circles (the pressure at $z<z_{\textrm{pr}}$ is fixed to be uniform in the 1D simulation). The profiles are normalized as $T_{8}\equiv T/10^{8}\,\textrm{K}$, $c_{s,8}\equiv c_{s}/10^{8}\,\textrm{cm}\,\textrm{s}^{-1}$ $p_{21}\equiv p/10^{21}\,\textrm{erg}\,\textrm{cm}^{-3}$, and $\rho_{4}\equiv\rho/10^{4}\,\textrm{g}\,\textrm{cm}^{-3}$.
}
\label{fig:FlashProf}
\end{figure}

The early evolution of the collision of two $0.64\,M_{\odot}$ CO WDs approaching with a Keplerian velocity of $v_0=2.3\times10^{3}\,\textrm{km}\,\textrm{s}^{-1}$ and with a zero impact parameter is shown in Figures~\ref{fig:Flash} (density map) and \ref{fig:FlashProf} (density, pressure, speed of sound, and temperature profiles along the axis of symmetry, $x=0$). The Figures correspond to $t=0.2\,\textrm{s}\simeq0.1v_{0}/g_{0}$ ($g_{0}\simeq1.2\times 10^8\,\textrm{cm}\,\textrm{s}^{-2}$). This collision was calculated by \citet{Kushnir2013} using high resolution 2D FLASH4.0 simulations with nuclear burning \citep[Eulerian, adaptive mesh refinement, 19 isotope alpha-chain reaction network,][]{dubey2009flash,Timmes1999}. The system of equations is closed with the {\it Helmholtz} equation of state \citep{Timmes2000} and a multipole gravity solver. Initially the CO WDs are at contact with free fall velocities. The structure of each CO WD is obtained from an isothermal stellar model\footnote{http://cococubed.asu.edu/code\_pages/adiabatic\_white\_dwarf.shtml} at $T=10^7$~K and with a uniform composition of $50\%$ Carbon and $50\%$ Oxygen by mass. Since the {\it Helmholtz} equation of state assumes complete ionization, the initial profile is not reliable for very low densities $\rho\simlt10^{4}\,\textrm{g}\,\textrm{cm}^{-3}$. For simplicity, we assume that the density profile at the lowest densities is a power law density profile $\rho\propto d^{-\omega}$, where $d$ is the distance from the stellar edge and $\omega\simeq-2.19$ determined by a fit to the profile from the stellar model in the vicinity of $\rho=10^{4}\,\textrm{g}\,\textrm{cm}^{-3}$. The exact profile at low densities has a small influence on our results, as discussed below.

This particular problem has a cylindrical symmetry and the position is described by cylindrical coordinates $x$ (radius with respect to the axis of symmetry) and $z$ (distance from the plane parallel to the surfaces at contact). The fact that the stars are identical implies a mirror symmetry $\pm z$ allowing us to focus on one of the stars. We note that the main features that are described below are not restricted to this 2D scenario.

As can be seen, shocks are propagating into each of the identical stars (reaching $\zsh\approx 0.5\times10^7\,\textrm{cm}$ at $t=0.2\,\textrm{s}$) as evident by the jump in density, pressure, speed of sound, and temperature. The velocity of the shocks $\Vzsh\simeq \zsh/t\simeq 250\,\textrm{km}\,\textrm{s}^{-1}$ is much smaller than $v_{0}$. As we show in Section~\ref{sec:SES}, generally $v_0/\Vzsh\simgt7$ for stellar collisions. Several interesting features are apparent in these figures which are generic to early phases of collisions.

\paragraph{The evolution has planar symmetry at early times, $t\ll (R_{\star}/v_0)(v_{0}/\Vzsh)^{2}$, in the vicinity of the contact region, $z,x\ll R_{\star}$} This is evident in figure \ref{fig:Flash} and results from the fact that the (cylindrical) radius of the contact region grows as $x_{\rm cont}\propto (R_{\star}v_{0}t)^{1/2}$ while the depth of the shocked region grows as $\zsh\propto \Vzsh t$ implying that at early times the shocked region is a thin disk with diverging aspect ratio $x_{\rm cont}/\zsh\propto [(R_{\star}/\Vzsh t)(v_0/\Vzsh)]^{1/2}\gg 1$. A quantitative illustration of the planar symmetry is provided in Figure~\ref{fig:FlashProf}, where the hydrodynamic profiles from a 1D planner numerical scheme, described in Sections~\ref{sec:1D} and~\ref{sec:1Ddetails}, are shown to agree (to better than $5\%$) with the profiles along the axis of symmetry. The initial CO WD density of the 1D model equals to the density on the axis of symmetry in the 2D model, and the initial velocity is the free fall velocity. The gravitational field is mimicked by an adjustable acceleration, which is constant in time and space. We choose to apply the surface acceleration of $g_{0}$, for which the close agreement between the two codes is found. However, since we are interested in the early evolution $t \ll v_{0}/g_{0}$, our results are not sensitive to the exact value of the applied acceleration. It is clear from Figure~\ref{fig:FlashProf} that it is difficult to infer the behavior of the flow variables near the contact surface directly from the 2D simulation because of the limited resolution.

\paragraph{The pressure is roughly uniform between the two shocks} This is a common feature of colliding mediums and is due to the short sound crossing time $\zsh/c_s$ compared to the evolution time $t$ which allows the pressure to be evenly distributed. This, in turn, is true due to the fact that the sound speed is generically faster than the velocity in which the shock moves with respect to the shocked fluid in the downstream ($z<\zsh$). In the scenario considered here the speed of sound is even larger at smaller $z$ as shown in Figure~\ref{fig:FlashProf} and explained below.

\paragraph{The speed of sound is diverging towards the contact surface} Given that the pressure is nearly uniform, this is directly related to the fact that the density is decreasing. The reason that the density is decreasing toward the contact surface is that these mass elements are near the surface of the star where the (pre-shocked) density is approaching zero. As we next show, while each element is adiabatically compressed by the flow, the compression is not sufficient to compensate for its initial low value.

\paragraph{The temperature reaches a finite value towards the contact surface} The vanishing density and the finite pressure near the contact surface imply that sufficiently close to the surface, the pressure is dominated by radiation, and therefore the temperature is given by $T\propto p^{1/4}$. As we show in Section~\ref{sec:NoBurn}, this temperature is generally too small for any appreciable nuclear burning at early times.

\subsection{The speed of sound generally diverges at the contact region}
\label{sec:general}

Consider a mass element which is in the vicinity of the contact region. It is useful to work with the column density
\begin{equation}
m=\int_0^z \rho dz
\end{equation}
as a Lagrangian coordinate, where $m=0$ at the contact surface. The pressure of the element at the current time $t$ is equal to the value of the pressure $p(t)$ throughout the shocked region where it is approximately uniform. Since the shock velocity at early time in the upstream frame is approximately constant (it is roughly $v_0$ because for stellar collision $v_0/\Vzsh\simgt7$, as we show in Section~\ref{sec:SES}), the pressure that the mass element had immediately after it was shocked is proportional to its pre-shocked density $p_{0,\rm sh}(m)\propto \rho(t=0,m)\equiv \rho_0(m)$. The increase in pressure since that time is $p(t)/p_{0,\rm sh}(m)\propto p(t)/\rho_0(m)$. Assuming that the adiabatic compression of the mass element behind the shock can be approximately described with an effective adiabatic index $\gamma$, the density and the speed of sound of the mass element at the current time are given by
\begin{eqnarray}\label{eq:RhoCs}
\rho(t,m)&\simeq& \rho_0(m)\left(\frac{p(t)}{p_{0,\rm sh}(m)}\right)^{1/\gamma}\propto p(t)^{1/\gamma}\rho_0(m)^{(\gamma-1)/\gamma}, \nonumber \\
c_{s}^{2}(t,m)&\propto& \frac{p(t)}{\rho(t,m)}\propto \left(\frac{p(t)}{\rho_0(m)}\right)^{\frac{\gamma-1}{\gamma}}.
\end{eqnarray}
Since $(\gamma-1)/\gamma>0$ (for fully ionized plasma $4/3<\gamma<5/3$), the temperature diverges at the contact surface, $m\rightarrow0$, where the pre-shocked density was vanishing.  We emphasize that the arguments leading to Eqeuation~\eqref{eq:RhoCs} do not depend on the assumptions that the stars have equal masses, that the impact parameter is zero or on the specific density profile. Note that the diverging speed of sound makes the approximation of a uniform pressure more accurate at the contact region. Since the pressure is radiation dominated in the vicinity of the contact region, the temperature there, $T_{c}(t)$, can be estimated as
\begin{eqnarray}\label{eq:Tc}
\frac{a}{3}T_{c}^{4}(t)&\simeq& p(t)\simeq\frac{2}{\gamma+1}\rho_{0}(t)v_{0}^{2}\Rightarrow \\
T_{c}(t)&\simeq&10^{9}\left(\frac{\rho_{0}(t)}{4\times10^{4}\,\textrm{g}\,
\textrm{cm}^{-3}}\right)^{1/4}\left(\frac{v_{0}}{3\times10^{3}\,\textrm{km}\,\textrm{s}^{-1}}\right)^{1/2}\textrm{K}\nonumber,
\end{eqnarray}
where $a$ is the black body radiation constant and $\rho_{0}(t)$ is the pre-shocked density of the element being shocked at the time $t$.

In the simple case that the initial density is a power law $\rho(t=0,z)=Kz^{-\omega}\propto m^{\omega/(\omega-1)}$ (see section \ref{sec:SES} for a detailed discussion), the pressure grows with time as $p(t)\propto t^{-\omega}$. The density and the speed of sound profile near the contact surface scale as
\begin{eqnarray}\label{eq:RhoCsPowerlaw}
\rho(t,m)&\propto& t^{-\omega/\gamma}m^{\frac{\omega(\gamma-1)}{\gamma(\omega-1)}}, \nonumber \\
c_{s}^{2}(t,m)&\propto& t^{-\frac{\omega(\gamma-1)}{\gamma}}m^{\frac{-\omega(\gamma-1)}{\gamma(\omega-1)}}.
\end{eqnarray}

The finite (nonzero) values of the density and the speed of sound at $z=0$ which were obtained in the 2D simulation presented in Figure \ref{fig:FlashProf} are the result of the finite resolution of that simulation. For higher resolution, the speed of sound (density) at $z=0$ increases (decreases). The diverging speed of sound implies that pressure can be quickly equilibrated in the contact plain and material is ejected in a thin layer parallel to the contact surface. The study of this ``jet'' is beyond the scope of this paper.

\subsection{A 1D numerical scheme that allows accurate calculations}
\label{sec:1D}

The planar geometry which is valid at early times allow the flow to be accurately solved for using a very high resolution 1D code. In particular, a fully Lagrangian code can be applied which allows the possible ignition of detonation process to be studied \citep{Kushnir2013, Kushnir2013b}. However, the compression of the elements close to the contact surface, combined with the fact that the speed of sound is diverging, significantly limits the time step allowed by the Currant condition. One way around this is to use Eulerian schemes for the flow near the contact surface, and restrict the size of the cells there. The disadvantage of this method is the numerical ``smearing'' of the flow variables near the contact surface on the scale of the cell size $\Delta z$ (which, for example, limits the maximal temperature there). Here we describe a 1D Lagrangian scheme, which allows to overcome the Currant time-step condition constrain without limiting the size of the innermost cells.

As discussed above, the mass elements near the contact surface equalize their pressure efficiently due to the high speed of sound. The scheme uses this fact by approximating the pressure of a chosen Lagrangian region, $0<m<m_{\rm pr}$, in the vicinity of the contact surface $m=0$, to be strictly uniform. In the outer regions, $m>m_{\rm pr}$, the hydrodynamic equations are solved as usual. The evolution of the hydrodynamic profile in the region $0<m<m_{\rm pr}$ is completely determined by the evolution of the value of the pressure in the region $p_{\rm pr}(t)$. Indeed, the density of each mass element depends on $p_{\rm pr}(t)$, and the position of each element can be expressed as $z(m)=\int dm \rho^{-1}$. The pressure $p_{\rm pr}(t)$ is determined by expressing the position of the edge of the uniform pressure region (which is determined similarly as in the outer region) in terms of the profiles in the region as
\begin{equation}\label{eq:zdotpdot}
z(m_{\rm pr})=\int_0^{m_{\rm pr}} dm \rho^{-1}.
\end{equation}
A discretized numerical scheme which implements such a region is described in Section~\ref{sec:1Ddetails}. In the numerical calculations presented below, this scheme was used. Nuclear burning can be added in a straight forward way and is described in \citet{Kushnir2013b}.

The choice of the (Lagrangian) position of the boundary $m_{\rm pr}$ is updated throughout the simulation to account for the growing region of nearly uniform pressure. In practice, a cell is added to the uniform pressure region if its Currant time-step becomes smaller than an appropriate threshold (as compared to outer cells) and convergence is verified by changing the threshold. The validity of this code is confirmed below where it is compared to the exact solutions which are obtained in the case of an ideal gas and power-law density profiles (Figure~\ref{fig:Profiles}).

\subsection{Lack of significant nuclear burning}
\label{sec:NoBurn}

The highest temperature in the vicinity of the contact surface are still too low to allow a significant nuclear burning. In particular, the ignition of a detonation is postponed to later stages of the evolution. This is a general feature of all CO WDs collisions. In the extreme case of $1\,M_{\odot}-1\,M_{\odot}$ collision, where $v_{0}\simeq3.5\times10^{3}\,\textrm{km}\,\textrm{s}^{-1}$, $g_{0}\simeq4.5\times 10^8\,\textrm{cm}\,\textrm{s}^{-2}$, and $t_{0}\simeq0.78\,\textrm{s}$, the upstream density at $0.1t_{0}$ is $\simeq2\times10^{4}\,\textrm{g}\,\textrm{cm}^{-3}$, leading to a temperature of $\simeq9.2\times10^{8}\,\textrm{K}$ (see Equation~\eqref{eq:Tc}) in the vicinity of the contact surface, which is too low for any significant nuclear burning. Note that this estimate derived under the condition $v_{0}/\Vzsh\gg1$, which we show below to generally hold for stellar collisions.
 % ---------------------------- End  of sec 2 ----------------------------

% -----------------------------------------------------------------------
% --------------------- Sec 3: n=1, omega<0 ---------------
% -----------------------------------------------------------------------
\section{Solution in the simplified case of an ideal gas and a power law density profile}
\label{sec:SES}

We showed in Section~\ref{sec:ContactProp} that at early times, where the velocity of the approaching stars is roughly constant, the problem has planar geometry. At these times the pressure in the pre-shocked region is negligible,  since close to the surface of the star the speed of sound decreases significantly and is much smaller than $v_0$, which is larger or equal to the free fall velocity and comparable to the typical speed of sound in the star. By adopting the following approximations, the problem is significantly simplified and allows an exact solution:
\begin{itemize}
\item The equation of state is that of an ideal gas, with an adiabatic index $\gamma$. This is not strictly correct for the shocked region, as near the shock the equation of state is close to that of an ideal gas ($\gamma=5/3$), while near the contact surface the pressure is radiation dominated ($\gamma=4/3$).
\item The density distribution has a power law dependence on the distance from the contact surface, \begin{equation}\rho(t=0,z)=Kz^{-\omega}.\end{equation}  While this is not strictly correct near the surface of WDs, by choosing appropriate values for  $\omega=-d\log\rho/d\log z$, the obtained solutions approximate the profiles at any given region. This approximation is also useful in collisions of other type of stars, with concrete examples including a radiative envelope with $\omega=-3$  or efficiently convective envelope (or degeneracy pressure) with $\omega=-3/2$ \citep{Chandrabook}.
\end{itemize}
The solution describes the position of the shock wave, $\zsh(t)$ and the hydrodynamical profiles between $z=0$ and $z=\zsh(t)$ at any given time $t$ after contact. In particular, we show that $v_{0}/\Vzsh\gtrsim7$ for all $\gamma$ and $\omega$ values that represent stellar collisions.

Since there are only three dimensional variables in the problem, $v_{0}$, $K$, and $t$, it is reasonable to assume that any dimensional quantity is given by the appropriate combination of these (up to a dimensionless multiplication factor) and the solution is self-similar. The profiles depend on the non-dimensional parameters $\omega$, $\gamma$ and normalized location $z/\zsh(t)$. The self-similarity allows the hydrodynamic partial equations to be reduced to an ordinary differential equation which can be easily (numerically) solved. The self-similarity solution is derived in Section~\ref{sec:SESexact}. The equations are solved in the frame of the upstream fluid, where the problem is equivalent to a piston moving into stationary fluid, and the results are transformed back to the laboratory frame of the stellar collisions. By comparing the obtained solutions to the results of direct 1D simulations of the same problem, the validity of the exact solutions and of the 1D numerical scheme is validated. In addition to the exact solution, a simplified analytic model is derived by approximating the pressure to be exactly uniform throughout the shocked region and is shown to provide an excellent approximation (Section~\ref{sec:SESexactSimple}). In Section~\ref{sec:self_similar_Appendix}, the solutions of the piston problem for planar, cylindrical and spherical symmetries, for all values of the density power law index $\omega$ are presented and compared.

\subsection{Exact self similar solution}
\label{sec:SESexact}
We solve the problem in the upstream frame, where it becomes identical to a planar piston moving with velocity $v_{0}$ into a stationary, cold fluid. The position of a fluid element located at a distance $z$ from the piston ($=$ contact surface) is given in this frame by $r=z+v_0t$ where the initial position of the piston is at $r=0$. The initial density profile is $\rho(t=0,r)=Kr^{-\omega}$. The position of the shock in this frame is given by $\Rsh(t)=\zsh(t)+v_0t$ and its velocity is $\Vsh(t)=\Vzsh(t)+v_{0}$. The local fluid velocity in this frame is $u=u_z+v_0$. The notations $r$ and $\Rsh$ are used to be consistent with the discussion in Section~\ref{sec:self_similar_Appendix} which includes cylindrical and spherical coordinate systems.

Before solving the hydrodynamic equations, we note that the velocity of the shock is exactly constant. Most simply, by using dimensional analysis, $\Rsh$ can only be constructed from $v_{0}t$ (we discuss the validity of the dimensional analysis in Section~\ref{sec:all}). Note that the shock cannot decelerate due to the zero velocity boundary condition at the piston and it cannot accelerate in the increasing density profile due to the limited energy budget at any given time (more details are given in Section~\ref{sec:all}).

Using dimensional arguments, it is possible to show that in the case where the flow is independent of any characteristic length scale, the flow fields must be of a self-similar form \citep[e.g.][]{ZeldoBook,Waxman2010} which we choose to be
\begin{equation}\label{eq:ss_scaling}
u=\Vsh\xi U(\xi),~c_{s}=\Vsh\xi C(\xi),~\rho=K\Rsh^{-\omega} G(\xi),
\end{equation}
where $u(r,t)$, $c_{s}(r,t)$, and $\rho(r,t)$ are the fluid velocity, sound speed, and density, respectively, and
\begin{equation}\label{eq:Rdot}
\xi(r,t)=r/\Rsh(t)=r/(\Vsh t)
\end{equation}
is the similarity parameter which is unity at the shock position. Note that the density scales like $\Rsh^{-\omega}$ since (for strong shocks) the density just behind the shock wave is a constant factor, $(\gamma+1)/(\gamma-1)$, times the pre-shocked density just ahead of the shock which is given by $K\Rsh^{-\omega}$.  The value of $\xi$ at the piston's position, $r_p=v_0t$, is given by
\begin{equation}
\xi_p=v_0/\Vsh.
\end{equation}
Finally, the pressure is
given by $p=\rho c^{2}/\gamma=K\Rsh^{-\omega} P(\xi)\dot{R}^2/\gamma$, where $P(\xi)=G(\xi)\xi^{2}C^{2}(\xi)$.

Using Equations~\eqref{eq:ss_scaling} and~\eqref{eq:Rdot} , the hydrodynamic equations, Equations~(\ref{eq:hydro_eq}) (with $n=1$ representing the planar case), can be expressed as a single ordinary differential equation, Equation~\eqref{eq:dUdC},
\begin{equation}\label{eq:dUdC_s}
\frac{dU}{dC}=\frac{\Delta_{1}(U,C)}{\Delta_{2}(U,C)},
\nonumber
\end{equation}
and one quadrature, Equation~(\ref{eq:quadrature}),
\begin{equation}\label{eq:quadrature_s}
\frac{d\ln\xi}{dU}=\frac{\Delta(U,C)}{\Delta_{1}(U,C)}\qquad {\rm
or} \qquad \frac{d\ln\xi}{dC}=\frac{\Delta(U,C)}{\Delta_{2}(U,C)},
\nonumber
\end{equation}
where $\Delta$, $\Delta_1$, and $\Delta_2$ are given by
\begin{eqnarray}\label{eq:deltas_s}
\Delta&=&C^{2}-f^{2}, \nonumber \\
\Delta_{1}&=&Uf^2-C^{2}\left(U-\frac{\omega}{\gamma}\right), \nonumber \\
\Delta_{2}&=&C\left[f^2-C^{2}+\frac{(\gamma-1)\omega}{2\gamma}\frac{C^{2}}{f}\right],
\end{eqnarray}
where
\begin{equation}
f=1-U.
\end{equation}

The boundary condition at the piston position is $u(t,r_p)=v_0$ which can be expressed as
\begin{equation}\label{eq:PistonBC}
U(\xi_p)=1.
\end{equation}
The Rankine--Hugoniot relations at the shock front determine the boundary conditions for the self-similar solutions to be \citep[e.g.][]{ZeldoBook}
\begin{equation}\label{eq:shock_boundary}
    U(1)=\frac{2}{\gamma+1},\quad C(1)=\frac{\sqrt{2\gamma(\gamma-1)}}{\gamma+1}, \quad G(1)=\frac{\gamma+1}{\gamma-1}.
\end{equation}

As illustrated here and in Section~\ref{sec:self_similar_Appendix} \citep[see also][]{Guderley42,MTV,Waxman93}, many of the properties of the self-similar flows may be inferred by analyzing the contours in the $(U,C)$-plane determined by Equation~(\ref{eq:dUdC}). Numerical integration of Equations~(\ref{eq:dUdC}) shows that for solutions starting at the strong shock point, Equations~(\ref{eq:shock_boundary}), $C$ diverges close to the piston as $U$ approaches 1 (see Figure~\ref{fig:UC} for a representative case), implying that the speed of sound is diverging near the piston as expected. We next analyze the behavior of the solution near the piston point $(U,C)=(1,\infty)$. Equation~(\ref{eq:dUdC}) is given, to leading order in $f$, by
\begin{equation}\label{eq:fC near edge}
\frac{d\ln f}{d\ln C}=\nu\frac{f}{C}\Rightarrow f\propto C^{\nu},
\end{equation}
where
\begin{equation}\label{eq:nu defenition}
\nu=\frac{2\left(\gamma-\omega\right)}{\omega\left(\gamma-1\right)}<0.
\end{equation}
The quadrature, Equation~(\ref{eq:quadrature}), gives to leading order in $f$
\begin{equation}\label{eq:f_xi_1}
    f=\left(1-\frac{\omega}{\gamma}\right)\ln\left(\frac{\xi}{\xi_{p}}\right).
\end{equation}
Using these results and Equation~(\ref{eq:G}) we find
\begin{equation}\label{eq:G1}
    G\propto f^{\omega(\gamma-1)/(\omega-\gamma)},
\end{equation}
and by using Equation~\eqref{eq:fC near edge}, we find that $P(\xi_{p})$ is finite (nonzero) implying that the pressure near the piston is not vanishing or diverging, as expected. We may now determine the dependence of the density and the speed of sound near the piston on mass and time. For a given mass element, we apply Equation~(\ref{eq:f_xi_1}) to its trajectory $\xi_0$ (which coincides with a $C_0$ characteristic),
\begin{equation}\label{eq:characteristics}
    \frac{dr}{dt}=u \Rightarrow \frac{d \ln \xi_{0}}{d \ln R}=-f(\xi_{0}),
\end{equation}
to get
\begin{equation}\label{eq:C0_1}
    f\propto\ln\left(\frac{\xi_0}{\xi_{p}}\right)\propto \Rsh^{-(1-\omega/\gamma)}\propto t^{-(1-\omega/\gamma)} .
\end{equation}
At a given time, the mass scales as $m\propto\rho\xi$. Using this result and Equation~\eqref{eq:C0_1} with Equation~\eqref{eq:G1} we find $\rho\propto t^{-\omega/\gamma}m^{\omega(\gamma-1)/\gamma(\omega-1)}$, and with Equation~\eqref{eq:fC near edge} we find $c_{s}^{2}\propto t^{-\omega(\gamma-1)/\gamma}m^{-\omega(\gamma-1)/\gamma(\omega-1)}$, the same dependence that was derived in Equation~\eqref{eq:RhoCsPowerlaw}. Note that since the pressure, density, and velocity are finite for the whole flow, the energy contained in the self-similar solution diverges as $t^{1-\omega}$ when $t\rightarrow\infty$, in accordance with the work done by the piston on the gas. This proves the consistency of the self-similar solution, as described with more details in Section~\ref{sec:all}.

\begin{figure}
\epsscale{1} \plotone{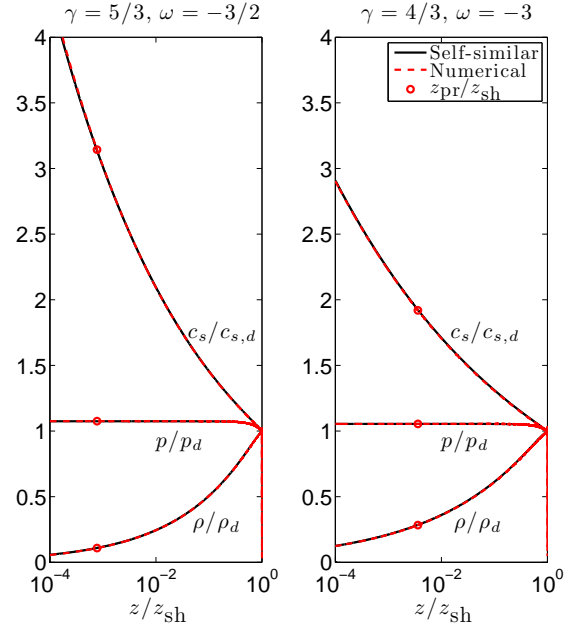} \caption{The self-similar profiles (speed of sound, pressure, and density) in the collision frame, normalized by their values immediately behind the shock front (denoted by a subscript $d$), are presented in black solid lines for two cases: $\gamma=5/3,\,\omega=-3/2$ (left) and $\gamma=4/3,\,\omega=-3$ (right). While the pressure in the shocked region is roughly uniform, the speed of sound (density) diverges (vanishes) near the contact surface. The profiles of the flow variables, from a numerical calculation, at the time the shock arrived to the edge of the computational grid, are presented in red dashed lines. The numerical profiles agree with the self-similar profiles to better than $1\%$ in the range $z/\zsh>10^{-6}$. The outer positions of the uniform pressure region at that time, $z_{\textrm{pr}}/\zsh$, are marked with red circles. The transition between the uniform pressure region and the regular region is smooth, and the numerical solution agrees with the self-similar solution in both regions. The excellent agreement between the results validates both the self-similar solution and the numerical scheme.}
\label{fig:Profiles}
\end{figure}

The solutions in the collision frame can be expressed by the solutions in the upstream frame with the relation
\begin{equation}\label{eq:frame_change}
\frac{z}{\zsh}=\frac{r-v_0t}{\Rsh-v_0t}=\frac{\xi-\xi_p}{1-\xi_p}.
\end{equation}
Note that in the collision frame the self-similar solution describes the whole space between the shock and the contact surface ($0\le z/\zsh \le 1$). The density, speed of sound and pressure are presented for two cases ($\gamma=5/3,\,\omega=-3/2$ and $\gamma=4/3,\,\omega=-3$) in Figure~\ref{fig:Profiles}. While the pressure in the shocked region is roughly uniform, the speed of sound diverges near the contact surface and the density vanishes there.

\begin{figure}
\epsscale{1} \plotone{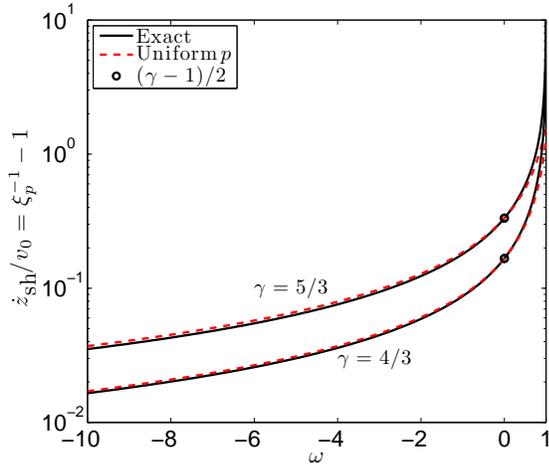} \caption{The shock velocities in the collision frame (normalized by $v_{0}$) as function of $\omega$ for $\gamma=4/3,5/3$ are shown in black. For cases relevant to stellar collisions ($\omega\le-3/2$), the shock velocity is a small fraction of $v_{0}$. This fraction increases with $\omega$, up to a value of $(\gamma-1)/2$ for $\omega=0$ (the solution of the corresponding Riemann problem). The behavior for $\omega>0$ is discussed in Section~\ref{sec:all_omega_gt_n}. The simple analytic solution given by Equation~\eqref{eq:Simple_vzsh_v0}, which is derived under the assumption that the pressure is uniform, is shown in red. The analytic expression provides an excellent approximation to the exact solution and is accurate to better than $6\%$ for $4/3<\gamma<5/3$ and $-10<\omega<-1$.}
\label{fig:ShockVelocity}
\end{figure}

The (normalized) shock velocity,
\begin{equation}
\frac{\Vzsh}{v_0}=\frac{\Vsh-v_0}{v_0}=\xi_p^{-1}-1,
\end{equation}
is provided for some values of $\gamma$ and $\omega$ in Table~\ref{tbl:ShockVelocityValues} and in Figure~\ref{fig:ShockVelocity}. As can be seen, for cases relevant to stellar collisions ($\omega\le-3/2$), the shock velocity is a small fraction of $v_{0}$. This fraction increases with $\omega$, up to a value of $(\gamma-1)/2$ for $\omega=0$ (the solution of the corresponding Riemann problem). The behavior for $\omega>0$ is discussed in Section~\ref{sec:all_omega_gt_n}. An analytic approximate expression for this velocity, Equation~\eqref{eq:Simple_vzsh_v0}, is derived below by approximating the pressure to be exactly uniform and is in excellent agreement with the exact result.

\begin{deluxetable}{ccc}
\tablecaption{Some values of $\Vzsh/v_{0}=\xi_{p}^{-1}-1$ as function of $\omega$ and $\gamma$ \label{tbl:ShockVelocityValues}}
\tablewidth{0pt} \tablehead{ \colhead{$\gamma$} & \colhead{$\omega=-3$} & \colhead{$\omega=-3/2$}}
\startdata $4/3$ & 0.0447  & 0.0704 \\
                $5/3$ & 0.0939  & 0.1462 \\
\enddata
\end{deluxetable}

\begin{figure}
\epsscale{1} \plotone{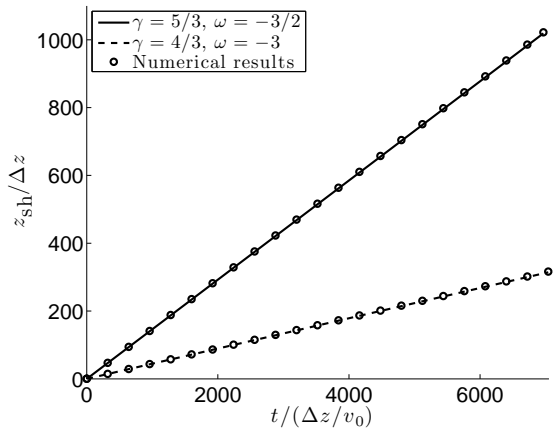} \caption{The shock trajectories in the collision frame are presented for two cases: $\gamma=5/3,\,\omega=-3/2$ (solid line) and $\gamma=4/3,\,\omega=-3$ (dashed line). The trajectories from a numerical calculation are presented in circles (the initial mesh consists of $8000$ cells with a uniform spacing, $\Delta z$). The agreement of the numerical trajectories with the self-similar trajectories is better than $0.5\%$.}.
\label{fig:ShockTrajectory}
\end{figure}

Next we compare the self-similar solution to the results of the direct $1$D numerical simulations described in section \ref{sec:1D} for the same ideal gas equation of state and density profiles. The initial mesh consists of $N$ cells with a uniform spacing, $\Delta z$. The initial pressure was chosen such that the outgoing shock wave is always strong. The shock trajectories for two representative cases, calculated with $N=8000$, are shown in Figure~\ref{fig:ShockTrajectory}. The agreement of the numerical trajectories with the self-similar trajectories is better than $0.5\%$. The profiles of the flow variables, at the time the shock arrived to the edge of the computational grid, are shown in Figure~\ref{fig:Profiles}. The numerical profiles agree with the self-similar profiles to better than $1\%$ in the range $\bar{\xi}>10^{-6}$. The outer positions of the uniform pressure region at that time, $z_{\textrm{pr}}/\zsh$, are marked with circles. The transition between the uniform pressure region and the regular region is smooth, and the numerical solution agrees with the self-similar solution in both regions. The excellent agreement between the results validates both the self-similar solution and the numerical scheme.

\subsection{Simple analytic solution}
\label{sec:SESexactSimple}

We next provide a simple analytic solution by using the fact that the pressure is nearly uniform. Assuming that the pressure is exactly uniform between the contact surface and the shock, we can use the same arguments that lead to equation \eqref{eq:RhoCs}, to express the velocity of the fluid element immediately at the downstream of the shock. The pressure $p_{\rm pr}(t)$ in the shocked region is given by its value in the immediate downstream and grows with time as $p_{\rm pr}(t)\propto \rho_0(t)\propto t^{-\omega}$. The density of a given mass element in the downstream grows with time as it is adiabatically compressed, $\rho(t,m)\propto p_{\rm pr}(t)^{1/\gamma}\propto t^{-\omega/\gamma}$. The size of each element thus shrinks according to $dx/dm=\rho^{-1}\propto t^{\omega/\gamma}$. This implies that the entire Lagrangian region between $0$ and $m$ scales in the same way $z(t,m)\propto t^{\omega/\gamma}$ and thus the fluid velocity of each element satisfies
\begin{equation}
u_z(m)=\frac{\partial z(t,m)}{\partial t}=\frac{\omega}{\gamma}\frac{z(t,m)}{t}.
\end{equation}
Applying this to the element which is immediately downstream of the shock we obtain the following equation:
\begin{equation}\label{eq:Simpleu_z}
u_z|_{\rm sh}=\frac{\omega}{\gamma}\Vzsh.
\end{equation}
By using the strong shock jump condition, $\Vzsh-u_z|_{\rm sh}=(\gamma-1)/(\gamma+1)(\Vzsh+v_0)$, we can solve for $\Vzsh$ in terms of $v_0$,
\begin{equation}\label{eq:Simple_vzsh_v0}
\frac{\Vzsh}{v_0}=\frac{\gamma(\gamma-1)}{2\gamma-\omega(\gamma+1)}.
\end{equation}
As can be seen in Figure~\ref{fig:ShockVelocity}, equation \eqref{eq:Simple_vzsh_v0} provides an excellent approximation to the exact solution and is accurate to better than $6\%$ for $4/3<\gamma<5/3$ and $-10<\omega<-1$. This expression does not capture the growing shock velocity (normalized by $v_0$) as $\omega$ approaches $0$ since the pressure is significantly non-uniform in the downstream region at these values. The hydrodynamic profiles under this approximation are exact power-laws with indexes as in Equation~\eqref{eq:RhoCsPowerlaw}, and amplitudes set by the shock jump conditions.
% ---------------------------- End  of sec 3 ----------------------------

% -----------------------------------------------------------------------
% --------------------- Sec 4: Discussion   ---------------
% -----------------------------------------------------------------------

\section{Summary and discussion}
\label{sec:Discussion}

The early phase of the hydrodynamic evolution following the collision of two stars is analyzed, focusing on the region near the contact region. It was shown in Section~\ref{sec:ContactProp} that the shocked region has a planar symmetry, a uniform pressure, and a diverging (vanishing) speed of sound (density) when approaching the contact surface (Equation~\eqref{eq:RhoCs}). The temperature reaches a finite value towards the contact surface (Equation~\eqref{eq:Tc}), since the vanishing density and the finite pressure near the contact surface imply that the pressure is dominated by radiation. We showed in Section~\ref{sec:NoBurn} that for all CO WDs collisions this temperature is generally too small for any appreciable nuclear burning at early times $t\lesssim0.1v_0/g_0$, before the velocity increases due to the gravitational acceleration. In particular, the ignition of a detonation is postponed to later stages of the evolution. This is tightly related to the fact that the shock moves very slowly in the collision frame compared to the fast approach speed $v_0$ (see Figure~\ref {fig:ShockVelocity} and Table~\ref{tbl:ShockVelocityValues}).

The divergence of the speed of sound has an important consequences for numerical studies of the stellar collisions. The numerical Currant condition will require a rapidly decreasing time step for higher resolutions, making convergence tests exceedingly expensive unless dedicated schemes are used. We described in Sections~\ref{sec:1D} and~\ref{sec:1Ddetails} a new 1D Lagrangian numerical scheme to achieve this.

We provided self-similar planar exact solutions for the simplified case of  a power-law density profile and an ideal equation of state in Section~\ref{sec:SESexact}. These solutions provide rough approximations that capture the main features of the flow and allow a general study as well as a detailed numerical verification test problem. Finally, we derived an approximate analytic expression for the shock velocity (Equation~\eqref{eq:Simple_vzsh_v0}) which is accurate to a few precent over a wide range of density profiles.

\acknowledgments We thank S. Dong, E. Waxman, and E. Livne for useful discussions. D.~K. gratefully acknowledges support from Martin A. and Helen Chooljian Founders' Circle. FLASH was in part developed by the DOE NNSA-ASC OASCR Flash Center at the University of Chicago. Computations were performed at PICSciE and IAS clusters.
% -------------------------- End of Discussion --------------------------

% -----------------------------------------------------------------------
% -------------------------------- Appendix -----------------------------
% -----------------------------------------------------------------------

\appendix

% -------------------------------- Appendix A --------------------------
\section{A. Self-similar piston driven flows}
\label{sec:self_similar_Appendix}

The problem of a piston propagating into a medium with a power law density profile which is studied in Section~\ref{sec:SESexact} (in the upstream frame) is the planar version of previous piston problems that were studied in cylindrical and spherical symmetries \citep[][and references therein]{Sedov45,Taylor46,SedovBook}. We found it timely to present a global picture of such self similar piston problems. In Section~\ref{sec:slfsim_eqns} we write down the hydrodynamic equations of the flow for planar, cylindrical and spherical symmetries ($n=1,2,3$, respectively) along with the resulting ODEs assuming self similarity.  The solutions for all geometries and all values of the density power law index $\omega$ are presented in the sections that follow. Particular emphasis is given to the non trivial transition to accelerating shocks at sufficiently declining densities, for which we derive new results and point out interesting similarities and differences with the strong explosion problem. A detailed discussion of the solutions for the planar case with growing density profiles ($\omega<0$), which are relevant for stellar collisions, is presented in Section~\ref{sec:SESexact}.

\subsection{A.1. The equations describing self-similar flows}
\label{sec:slfsim_eqns}

In the problems considered, a piston moves with a constant velocity $v_0$ into a medium with an ideal gas equation of state with an adiabatic index $\gamma$, and an initial power-law density profile
\begin{equation}
\rho(t=0,r)=Kr^{-\omega},
\end{equation}
where $r$ is the radial coordinate. The medium is assumed to have zero pressure initially and a strong shock propagates ahead of the piston. The equations describing the adiabatic 1D flow behind the shock are \citep[e.g.][]{LandauBook}
\begin{eqnarray}
\label{eq:hydro_eq}
(\partial_{t}+u\partial_{r})\ln\rho+ r^{-(n-1)}\partial_{r}(r^{n-1}u) &=& 0,
\nonumber \\
(\partial_{t}+u\partial_{r})u+\rho^{-1}\partial_{r}(\gamma^{-1}\rho c_{s}^{2}) &=&
0, \nonumber \\
(\partial_{t}+u\partial_{r})(c_{s}^{2}\rho^{1-\gamma}) &=& 0,
\end{eqnarray}
where $n=1,2,3$ are for planar, cylindrical, and spherical symmetry, respectively.

Self similar solutions are obtained by substituting Equations~(\ref{eq:ss_scaling}) and~(\ref{eq:Rdot}) in the hydrodynamic Equations~(\ref{eq:hydro_eq}), with a shock velocity scaling (the shock velocity is not constant in general),
\begin{equation}\label{eq:Rdot_delta}
    \Vsh=A\Rsh^\delta.
\end{equation}
The partial differential equations, Equations~\eqref{eq:hydro_eq}, are replaced with a single ordinary differential equation  \citep{ZeldoBook,Waxman2010},
\begin{equation}\label{eq:dUdC}
\frac{dU}{dC}=\frac{\Delta_{1}(U,C)}{\Delta_{2}(U,C)},
\end{equation}
and one quadrature
\begin{equation}\label{eq:quadrature}
\frac{d\ln\xi}{dU}=\frac{\Delta(U,C)}{\Delta_{1}(U,C)}\qquad {\rm
or} \qquad \frac{d\ln\xi}{dC}=\frac{\Delta(U,C)}{\Delta_{2}(U,C)}.
\end{equation}
The normalized density, $G$, is given implicitly by
\begin{equation}\label{eq:G}
(\xi C)^{-2(n-\omega)}|1-U|^{\lambda}G^{(\gamma-1)(n-\omega)+\lambda}\xi^{n\lambda}={\rm const}
\end{equation}
with
\begin{equation}\label{eq:lambda}
\lambda=\omega(\gamma-1)+2\delta.
\end{equation}
The functions $\Delta$, $\Delta_{1}$, and $\Delta_{2}$ are
\begin{eqnarray}\label{eq:deltas}
\Delta&=&C^{2}-(1-U)^{2}, \nonumber \\
\Delta_{1}&=&U(1-U)(1-U-\delta)-C^{2}\left(n U+\frac{2\delta-\omega}{\gamma}\right), \nonumber \\
\Delta_{2}&=&C\{(1-U)(1-U-\delta) \nonumber \\
&-&\frac{\gamma-1}{2}U\left[(n-1)(1-U)+\delta\right]-C^{2} \nonumber \\
&+&\frac{2\delta+\omega(\gamma-1)}{2\gamma}\frac{C^{2}}{1-U}\}.
\end{eqnarray}

\subsection{A.2. The self similarity assumption}
\label{sec:all}

There are two types of similarity solutions \citep[see e.g.][]{ZeldoBook}. Following \citet{KushnirGap}, solutions of the first-type may be defined as solutions that are valid over the entire $(r,t)$-plane (or the part of which where the flow takes place). Such solutions must satisfy the global conservations laws of mass, momentum, and energy, and hence the values of the similarity exponents of such solutions may be determined by dimensional considerations. Solutions of the second-type may be defined as solutions which describe only part of the flow. Such solutions should be required to allow the existence of a characteristic line, $\xi_c(\Rsh)$, along which the self-similar solution is matched to another solution, and to comply with the global conservation laws within the region of the $(r,t)$-plane described by the self-similar solution \citep[note that it is commonly accepted that the similarity exponents of a second-type solution are determined by the requirement that the solution passes through a singular point of the hydrodynamic equations, but this condition is not general enough, see][]{KushnirGap}.

Before solving the hydrodynamic equations, we can use simple arguments to derive some properties of the self-similar solutions. To begin with, the shock must propagate with a constant velocity ($\delta=0$) or accelerate ($\delta>0$), since if it is decelerating, the piston reaches it at some finite time. Next, let us assume that for the cases where the mass near the piston is finite  ($\omega<n$), the self-similar solution is valid everywhere between the piston and the shock (first-type solution). Below we show that this assumption results in a consistent solution. For $\omega\ge n$, the mass near the piston diverges and there are no consistent self similar solutions of the entire flow. We discuss second-type self-similar solutions for this regime in Section~\ref{sec:all_omega_gt_n}. In the first-type case, the shock must propagate at a constant velocity, as can be derived from a few arguments. Most simply, by using dimensional analysis, $\Rsh$ can only be constructed from $v_{0}t$. Another argument concerns the mass element adjacent to the piston. Such an element is part of the self-similar flow and therefore its normalized position, $\xi_{p}=v_{0} t/\Rsh(t)$, must be constant, implying that $\delta=0$. Moreover, since its position coincides with a $C_{0}$ characteristic of the self-similar solution, given by Equation~\eqref{eq:characteristics}, we must have $U(\xi_{p})=1$. By using Equation~(\ref{eq:ss_scaling}), we can derive the shock velocity, $\Vsh=v_{0}/\xi_{p}$, which also equals the constant $A$.

A more physical argument for the constant velocity of the shock can be made by considering the total energy of the flow. The energy contained in the self-similar solution is
\begin{eqnarray}\label{eq:E_ss general}
E_{s}(\Rsh)&=f(n)&\int_{\xi_{\rm p}\Rsh}^{\Rsh} dr r^{n-1}\left(\frac{1}{2}\rho u^2+\frac{1}{\gamma-1}p\right) \nonumber\\
&=&f(n)A^{2}\Rsh^{2\delta+n-\omega}K
\left[I_{k}(\xi_{p})+I_{i}(\xi_{p})\right],
\end{eqnarray}
with
\begin{equation}\label{eq:Ik def general}
I_{k}(\xi)=\int\limits_{\xi}^{1}d\xi'\xi'^{n+1}G\frac{1}{2}U^{2}, \quad
I_{i}(\xi)=\int\limits_{\xi}^{1}d\xi'\xi'^{n+1}G\frac{1}{\gamma(\gamma-1)}C^{2},
\end{equation}
and
\begin{equation}\label{eq:f def}
f(n)=
\begin{cases}
    {1} & \text{for $n=1$,} \\
    {2\pi} & \text{for $n=2$,} \\
    {4\pi} & \text{for $n=3$.} \
\end{cases}
\end{equation}
The $I_{k}$ and $I_{i}$ terms describe the kinetic and internal energy of the gas, respectively. Since at any given shock position $\Rsh$ the energy of the gas must be finite (nonzero), $I_{k}$ and $I_{i}$ cannot diverge and at least one of them is nonzero. Therefore, $E_s(\Rsh)$ diverges as $\Rsh^{2\delta+n-\omega}$ when $\Rsh\rightarrow\infty$. The energy of the gas is supplied from the work done on it by the piston,
\begin{equation}\label{eq:piston work}
W(\Rsh)\propto\int\limits_{0}^{t(\Rsh)}dtv_{0}(v_{0}t)^{n-1}\Rsh^{2\delta-\omega}P(\xi_{p})\propto P(\xi_{p})\Rsh^{2\delta+n/\alpha-\omega}.
\end{equation}
In order for the work done by the piston to diverge in accordance with the energy of the gas, we must have $\alpha=1$ ($\delta=0$). For the work done by the piston to be finite (nonzero), $P(\xi_{p})$ must be finite (nonzero), and therefore the pressure near the piston behaves as $p(t)\propto t^{-\omega}$. Note that if $\delta=0$ for $n\ge\omega$, then the energy of the gas is not increasing as $\Rsh\rightarrow\infty$, which is not physical given that the piston is performing work on the gas. This immediately shows that for $n\ge\omega$ the self-similar solution cannot be valid near the piston, and only second-type self-similar solutions are possible, with the possibility that the shock accelerates, $\delta>0$ (see Section~\ref{sec:all_omega_gt_n}).

It is straightforward to generalize Equation~\eqref{eq:RhoCsPowerlaw} to the general case $\omega<n$, by noting that $\rho(t=0,r)=Kr^{-\omega}\propto m^{\omega/(\omega-n)}$. The density and the speed of sound profiles near the contact scale as
\begin{eqnarray}\label{eq:RhoCsPowerlawAll}
\rho(t,m)&\propto& t^{-\omega/\gamma}m^{\frac{\omega(\gamma-1)}{\gamma(\omega-n)}}, \nonumber \\
c_{s}^{2}(t,m)&\propto& t^{-\frac{\omega(\gamma-1)}{\gamma}}m^{\frac{-\omega(\gamma-1)}{\gamma(\omega-n)}}.
\end{eqnarray}
These results can be verified directly from the asymptotic behavior near the piston, derived below. The density near the piston vanishes (diverges) and the speed of sound there diverges (vanishes) for $\omega<0$ ($\omega>0$). For $\omega=0$ both density and speed of sound are finite (non-zero) near the piston \citep[for $n=1$ this is a simple Riemann problem, while for $n=2,3$ the solutions were derived by][]{Sedov45,Taylor46,SedovBook}.

In Sections~\ref{sec:all_omega_lt_0}, \ref{sec:all_omega_eq_0}, and \ref{sec:all_omega_gt_0} we show for $\omega<0$, $\omega=0$, and $0<\omega<n$, respectively, that $I_{k}$ and $I_{i}$ are finite, such that $E_s(\Rsh)$ diverges as $\Rsh^{n-\omega}$ when $\Rsh\rightarrow\infty$, in accordance with the work done by the piston on the gas. This showes that the first-type self-similar solutions are consistent. The derived shock velocity is shown for some values of $\gamma$ and $\omega$ in Figures~\ref{fig:ShockVelocity} and~\ref{fig:DeltaOmega} for the $n=1$ and $n=2,3$ cases, respectively.

\subsection{A.3. Increasing density profile ($\omega<0$)}
\label{sec:all_omega_lt_0}

\begin{figure}
\epsscale{1} \plotone{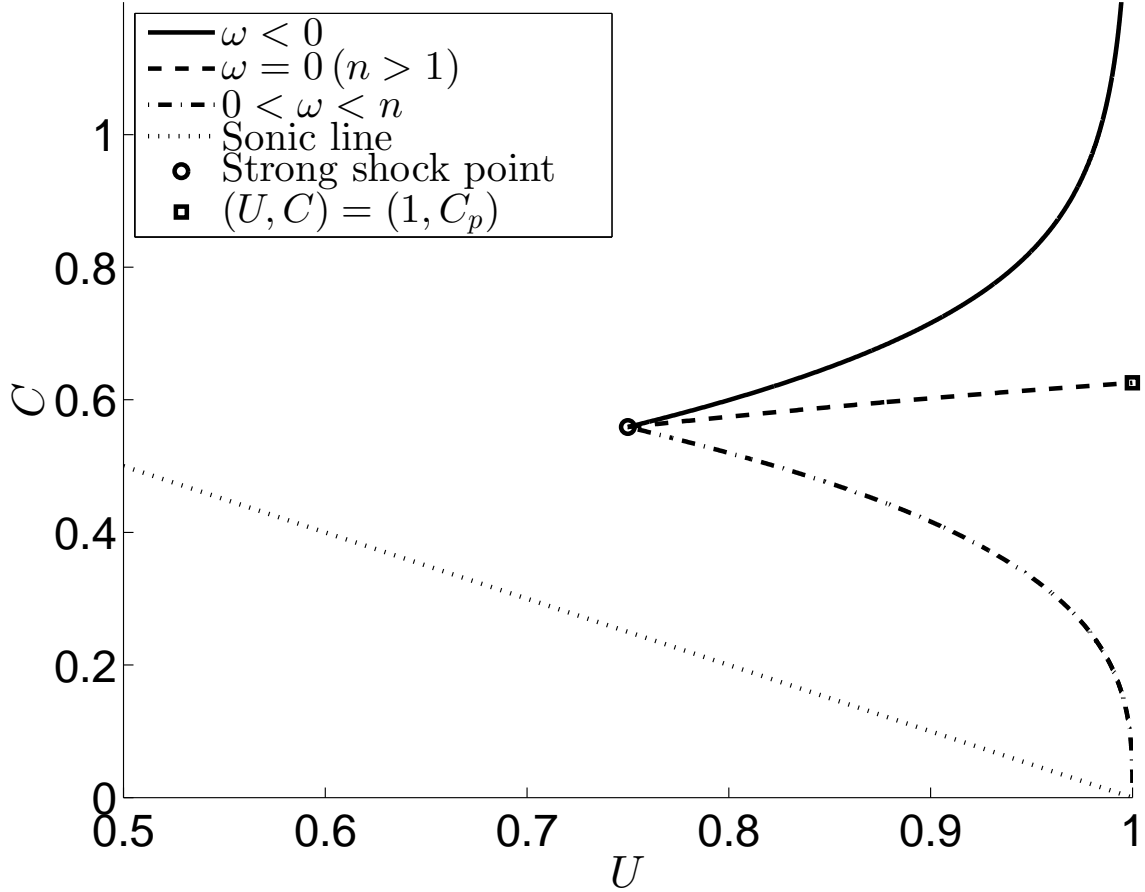} \caption{Different types of the $C(U)$ curves for the solutions of the piston problem, obtained for $\omega$ values in the regimes $\omega<0$ (solid), $\omega=0$ (dashed), and $0<\omega<n$ (dot-dashed). Also shown as a dotted line is the sonic line, $U+C=1$. The circle marks the strong shock point, Equations~(\ref{eq:shock_boundary}), and the square marks the point$(U,C)=(1,C_{p})$.}
\label{fig:UC}
\end{figure}

Numerical integrations of Equations~(\ref{eq:dUdC}) starting at the strong shock point, Equations~(\ref{eq:shock_boundary}), indicate that $C$ diverges as $U$ approaches 1 (see Figure~\ref{fig:UC} for a representative case). Analysis of the the behavior of the solution near $(U,C)=(1,\infty)$ shows that $f\propto C^{\nu}$, where
\begin{equation}\label{eq:nu defenition general}
\nu=\frac{2\gamma\left(n-\omega/\gamma\right)}{\omega\left(\gamma-1\right)}<0.
\end{equation}
The quadrature, Equation~(\ref{eq:quadrature}), gives to leading order in $f$
\begin{equation}\label{eq:f_xi_1 omega<0}
    f=\left(n-\frac{\omega}{\gamma}\right)\ln\left(\frac{\xi}{\xi_{p}}\right).
\end{equation}
Using these results and Equation~(\ref{eq:G}) we find
\begin{equation}\label{eq:G1 omega<0}
    G\propto f^{\omega(1-\gamma)/(n\gamma-\omega)}.
\end{equation}
Therefore, $I_{k}$ and $I_{i}$ are finite.

\subsection{A.4. Constant density profile ($\omega=0$)}
\label{sec:all_omega_eq_0}

For planner symmetry ($n=1$) this is a simple Riemann problem, and in what follows we consider the cylindrical and spherical ($n=2,3$, respectively) cases \citep{Sedov45,Taylor46,SedovBook}. Numerical integration of Equations~(\ref{eq:dUdC}) shows that for solutions starting at the strong shock point, $C$ reaches a finite (nonzero) value $C_{p}$ as $U$ approaches 1 (see Figure~\ref{fig:UC} for a representative case). Note that $(U,C)=(1,C_{p})$ is not a singular point of Equations~(\ref{eq:dUdC}), but nevertheless the integration ends there, as the boundary condition for the piston is $U(\xi_{p})=1$. We next analyze the the behavior of the solution near $(U,C)=(1,C_{p})$. To leading order in $f$, Equation~(\ref{eq:dUdC}) is given by
\begin{equation}\label{eq:fC near edge omega=0}
\frac{df}{dC}=-\frac{n}{C_{p}}\Rightarrow f=n\left(1-\frac{C}{C_{p}}\right).
\end{equation}
The quadrature, Equation~(\ref{eq:quadrature}), gives to leading order in $f$
\begin{equation}\label{eq:f_xi_1 omega=0}
    f=n\ln\left(\frac{\xi}{\xi_{p}}\right).
\end{equation}
Using these results and Equation~(\ref{eq:G}) we find that $G$ has some finite (nonzero) value near the piston. Therefore, $I_{k}$ and $I_{i}$ are finite.

\subsection{A.5. Moderately decreasing density profile such that the mass near the piston is finite ($0<\omega<n$)}
\label{sec:all_omega_gt_0}

Numerical integration of Equations~(\ref{eq:dUdC}) shows that for solutions starting at the strong shock point, $C$ reaches zero as $U$ approaches 1. We next analyze the the behavior of the solution near $(U,C)=(1,0)$. We consider the planar case seperately ($n=1$) since its analysis is somewhat different than that of the cylindrical and spherical cases ($n=2,3$).

\subsubsection{A.5.1 Planar symmetry ($n=1$)}
\label{sec:omega_gt_0_n=1}

Equation~(\ref{eq:dUdC}) is given, to leading order in $f$, by
\begin{equation}\label{eq:fC near edge omega>0, n=1}
\frac{d\ln f}{d\ln C}=-\frac{f^{2}-C^{2}\left(1-\frac{\omega}{\gamma}\right)}{f^{3}+\frac{\omega(\gamma-1)}{2\gamma}C^{2}}.
\end{equation}
Equation~\eqref{eq:fC near edge omega>0, n=1} implies that
\begin{equation}\label{eq:f on C near edge}
\lim_{f\to 0}\frac{d\ln f}{d\ln C}=\nu,
\end{equation}
where either $\nu=1$ or $\nu=2(\gamma-\omega)/\omega(\gamma-1)>1$. The quadrature, Equation~(\ref{eq:quadrature}), gives to leading order in $f$
\begin{equation}\label{eq:f_xi_1 omega>0, n=1}
    f=\theta\ln\left(\frac{\xi}{\xi_{p}}\right),\quad\theta=
    \left\{
      \begin{array}{ll}
        \left(1-\frac{\omega}{\gamma}\right), & \hbox{$\nu>1$ ,} \\
        \frac{(\gamma-1)}{\gamma+1}, & \hbox{$\nu=1$.}
      \end{array}
    \right.
\end{equation}
Using these results and Equation~(\ref{eq:G}) we find
\begin{equation}\label{eq:G1 omega>0, n=1}
G\propto f^{-\mu}, \quad \mu=
    \left\{
      \begin{array}{ll}
        (\gamma-1)\omega/(\gamma-\omega), & \hbox{$\nu>1$ ,} \\
        ((\gamma+1)\omega-2)/(\gamma-1), & \hbox{$\nu=1$.}
      \end{array}
    \right.
\end{equation}
Numerical integration of Equations~(\ref{eq:dUdC}) and~(\ref{eq:quadrature}) shows that solutions starting at the strong shock point approach $(U,C)=(1,0)$ along a $\nu>1$ curve. However, for both $\nu=1$ and $\nu>1$, $I_{k}$ and $I_{i}$ are finite.	

As can be seen in Figure~\ref{fig:ShockVelocity}, the ratio between the shock velocity and $v_{0}$ diverges in the limit where $\omega\rightarrow1$. We note that $P(\xi_{p})$ approaches zero in this limit.

\subsubsection{A.5.2. Cylindrical and spherical symmetries ($n=2,3$)}
\label{sec:omega_gt_0_n=2,3}

In this case, equation~(\ref{eq:dUdC}) is given, to leading order in $f$, by
\begin{equation}\label{eq:fC near edge omega>0, n=2,3}
\frac{d\ln f}{d\ln C}=-\frac{f^{2}-C^{2}\left(n-\frac{\omega}{\gamma}\right)}{-\frac{(\gamma-1)(n-1)}{2}f^{2}+\frac{\omega(\gamma-1)}{2\gamma}C^{2}}.
\end{equation}
The solution of Equation~(\ref{eq:fC near edge omega>0, n=2,3}) must also be of the form given by Equation~(\ref{eq:f on C near edge}). Assuming $f$ tends to 0 slower than $C$, i.e., $\nu<1$, leads to a contradiction since Equation~(\ref{eq:fC near edge omega>0, n=2,3}) gives $\nu=2/(\gamma-1)(n-1)$, which is larger than $1$ for $\gamma<3(2)$ and $n=2(3)$. Therefore, $\nu$ must satisfy $\nu\ge1$. For $\nu>1$, Equation~(\ref{eq:fC near edge omega>0, n=2,3}) gives
\begin{equation}\label{eq:f_C omega>0, n=2,3}
    \nu=\frac{2(n\gamma-\omega)}{\omega(\gamma-1)},
\end{equation}
which satisfies $\nu>1$ for $\omega<n$. For $\nu=1$, Equation~(\ref{eq:fC near edge omega>0, n=2,3}) gives
\begin{equation}\label{eq:f_C_nu1 omega>0, n=2,3}
    f^2=\frac{2n\gamma-(\gamma+1)\omega}{2\gamma-\gamma(\gamma-1)(n-1)} C^2.
\end{equation}
The solution of the quadrature, Equation~(\ref{eq:quadrature}), gives
\begin{equation}\label{eq:f_xi}
    f=\theta\ln\left(\frac{\xi}{\xi_{p}}\right),\quad\theta=
    \left\{
      \begin{array}{ll}
        \left(n-\frac{\omega}{\gamma}\right), & \hbox{$\nu>1$ ,} \\
        \frac{n(\gamma-1)}{\gamma+1}, & \hbox{$\nu=1$.}
      \end{array}
    \right.
\end{equation}
Using these results and Equation~(\ref{eq:G}) we find
\begin{equation}\label{eq:Geta}
    G\propto f^{-\mu}, \quad \mu=
    \left\{
      \begin{array}{ll}
        (\gamma-1)\omega/(n\gamma-\omega), & \hbox{$\nu>1$ ,} \\
        ((\gamma+1)\omega-2n)/n(\gamma-1), & \hbox{$\nu=1$.}
      \end{array}
    \right.
\end{equation}
Numerical integration of Equations~(\ref{eq:dUdC}) and~(\ref{eq:quadrature}) shows that solutions starting at the strong shock point approach $(U,C)=(1,0)$ along a $\nu>1$ curve. However, $\mu<1$ for both $\nu=1$ and $\nu>1$, and thus $I_{k}$ and $I_{i}$ are finite.

As can be seen in panel (d) of Figure~\ref{fig:DeltaOmega}, $\Vsh/v_{0}$ is some finite value (which depends on $\gamma$) at $\omega=2(3)$ for $n=2(3)$ (unlike the planner case, for which $\Vsh/v_{0}$ diverges as $\omega\rightarrow1$).

\subsection{A.6. Steeply decreasing density profile such that the mass near the piston diverges ($\omega\ge n$)}
\label{sec:all_omega_gt_n}

We already commented, based on the energy contained in the first-type self-similar solution, that for $\omega\ge n$ the self-similar solution cannot be valid near the piston, and only second-type self-similar solutions are possible. In what follows, we derive this directly from the properties of the self-similar solutions, assuming $\delta=0$.

Numerical integration of Equations~(\ref{eq:dUdC}) for $n=1$ shows that for solutions starting at the strong shock point, Equations~(\ref{eq:shock_boundary}), $C(U)$ crosses the sonic line ($U+C=1$) at a non-singular point (either $\Delta_{1}\ne$ and/or $\Delta_{2}\ne0$). Therefore, $\delta=0$ results in a non-physical solution. For $n=2,3$, the integration shows that in the range $n\le\omega\le\omega_{g}(\gamma,n)$ \citep[$\omega_g$ is increasing with $\gamma$, $\omega_{g}(1,n)=n$ and $\omega_{g}(5/3,n)\simeq 2.09(3.26)$ for $n=2(3)$, see][for the $n=3$ case]{Waxman93}, known as the ``gap'' region, $C$ reaches zero as $U$ approaches 1. For $\omega$ values above the ``gap'', $C(U)$ crosses the sonic line at a non-singular point. Thus, we only need to show that in the ``gap'', $\delta=0$ results in a non-physical solution. This is achieved by examining Equation~\eqref{eq:Geta}, which shows that $\mu>1$ for both $\nu=1$ and $\nu>1$, and therefore $I_{k}$ diverges in the limit $\xi\rightarrow\xi_{p}$ \citep[this was shown for $n=3$ in][]{KushnirGap}.

The second-type self-similar solutions were already found for the strong explosion problem, in which a large amount of energy is deposited within a small region at the center of an initially cold gas with an initial density $\rho=K r^{-\omega}$, for $3\le\omega\le\omega_{g}(\gamma,3)$  by \citet{KushnirGap} and for $\omega_{g}(\gamma,3)<\omega$ by \citet{Waxman93,Waxman2010}. The equations for the strong explosion problem and for the piston problem are identical for second-type self-similar solutions \cite[since $\delta$ is uniquely determined by the existence of such a solution,][]{KushnirGap}. Given that the self-similar part of the flow does not depend on the details of the non-self-similar part of the flow (one case involving an explosion, following which each mass element moves at an asymptotically constant velocity and the second case involving a piston moving at a constant velocity), these second-type self-similar solutions are also the solutions for the piston problem.

\begin{figure}
        \subfigure[]{
             \includegraphics[width=0.5\textwidth]{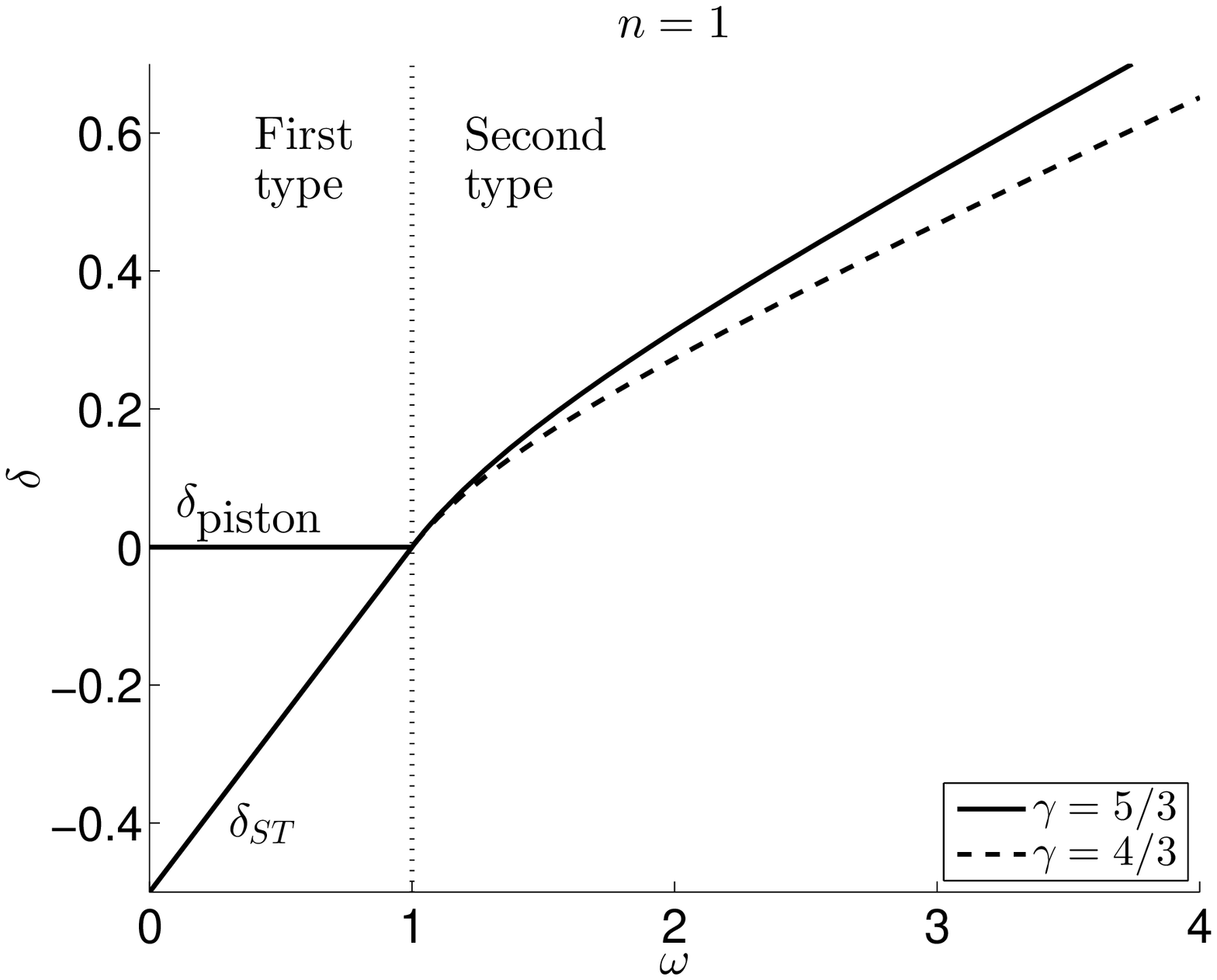}
        }
        \subfigure[]{
             \includegraphics[width=0.5\textwidth]{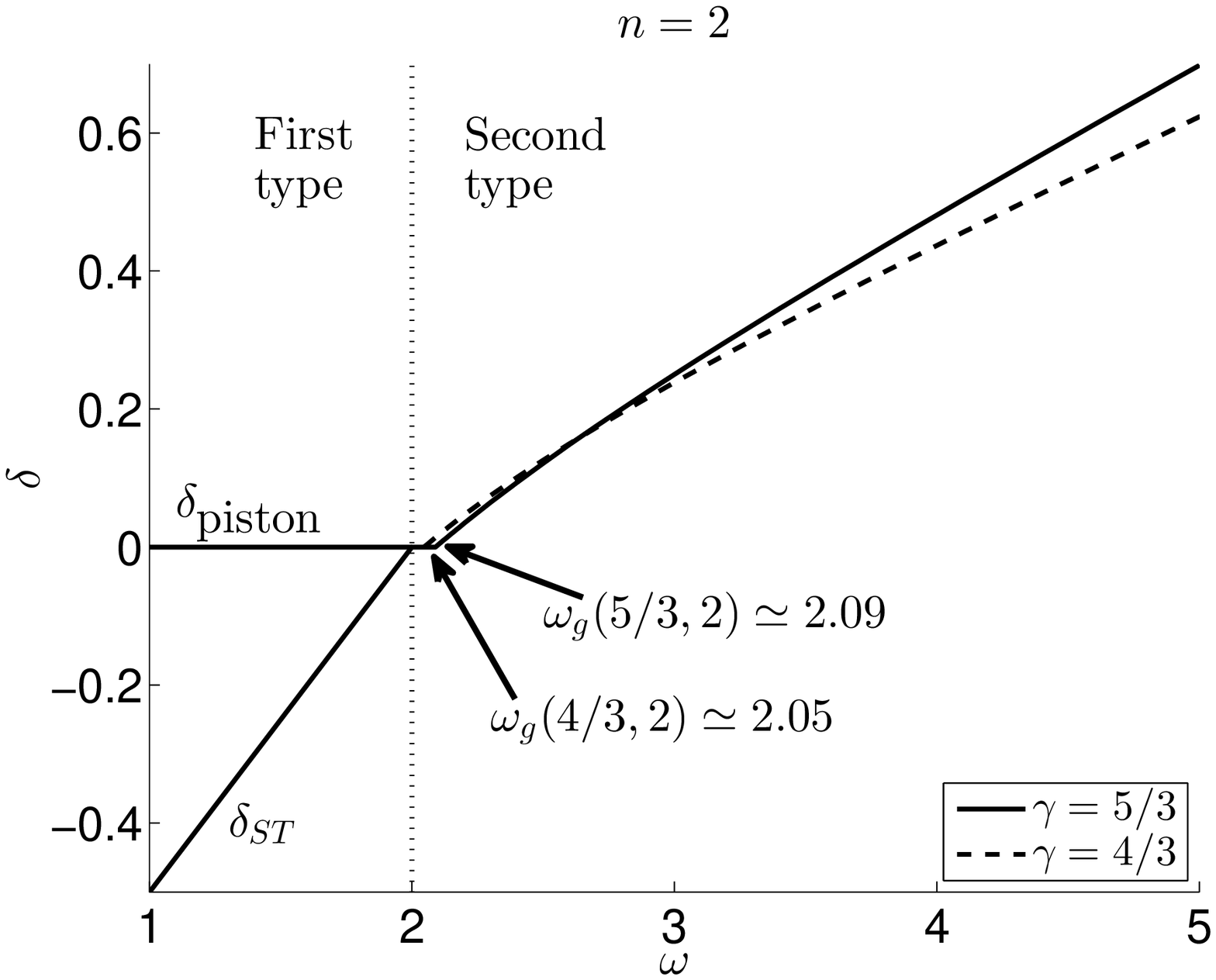}
        }
        \subfigure[]{
             \includegraphics[width=0.5\textwidth]{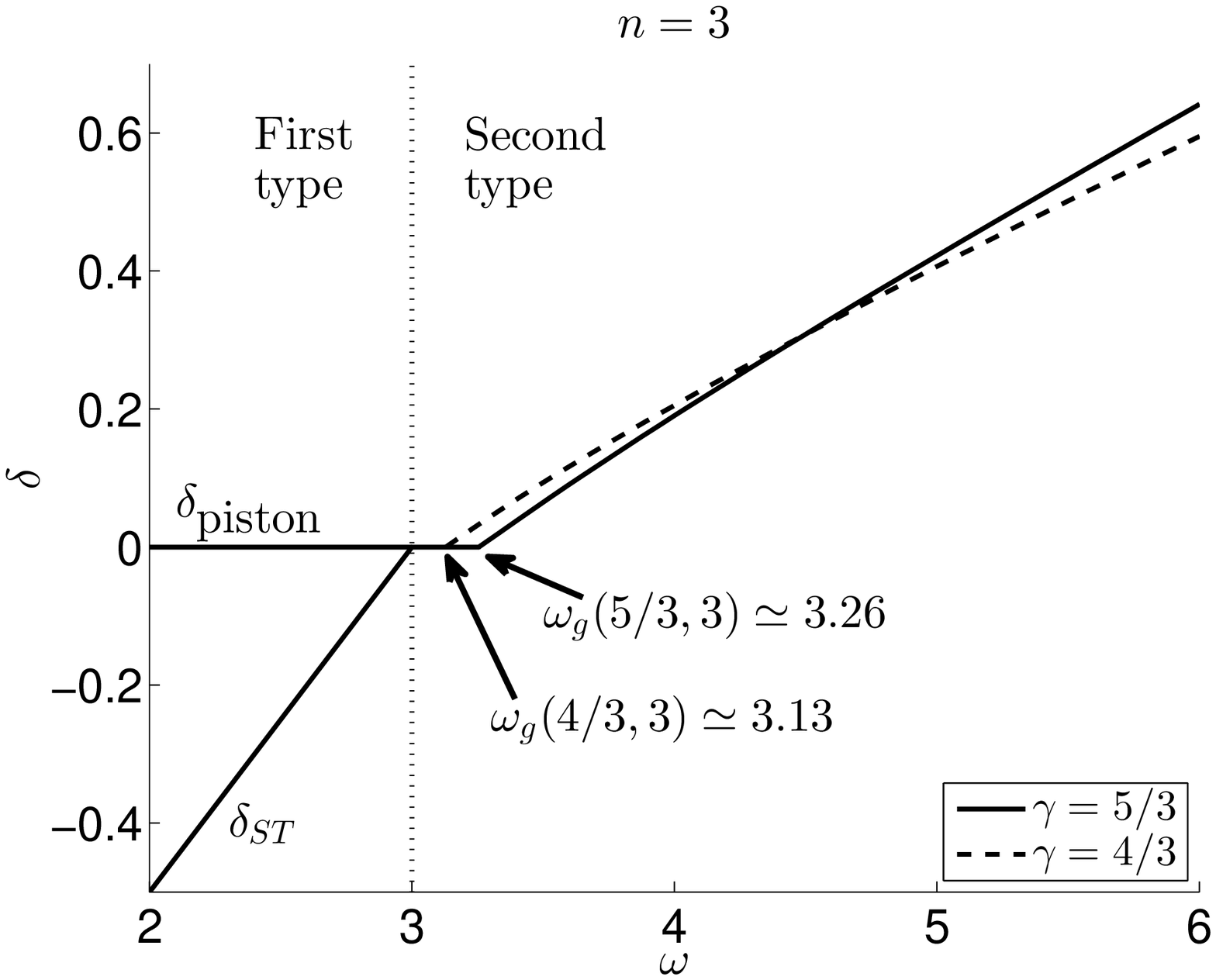}
        }
        \subfigure[]{
             \includegraphics[width=0.5\textwidth]{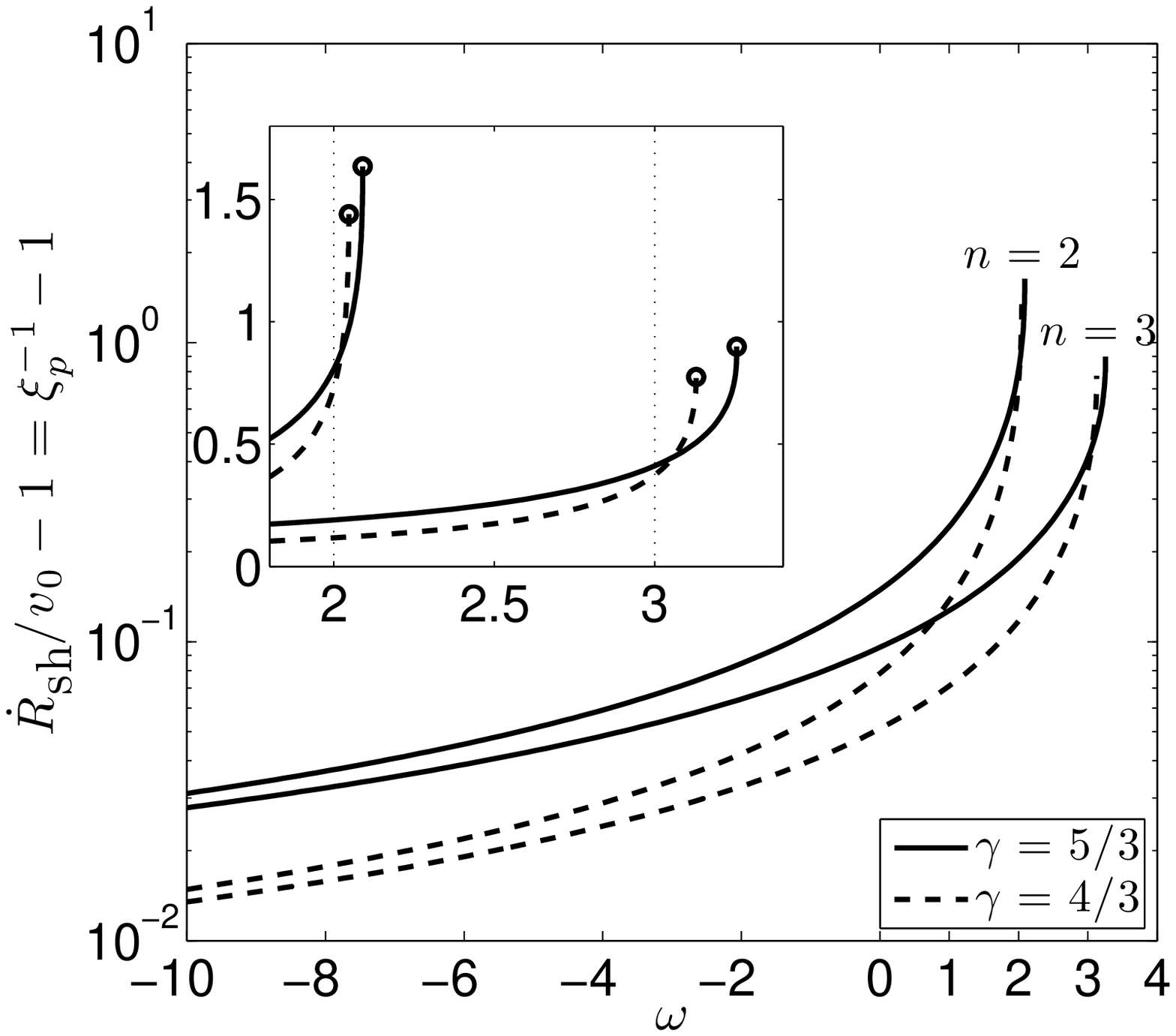}
        }
      \caption{Panels (a)-(c): the self-similar acceleration exponent
$\delta$ (where $\dot R_{\rm sh}\propto R_{\rm sh}^\delta$) as a function of the density profile index $\omega$ (where $\rho(t=0,r)\propto r^{-\omega}$) for $\gamma=5/3$ (solid line) and $\gamma=4/3$ (dashed line). Panel (a): planar symmetry ($n=1$). Panel (b): Cylindrical symmetry ($n=2$). Panel (c): Spherical symmetry ($n=3$). For the strong explosion problem and $\omega<n$, the self-similar exponent satisfies $\delta_{ST}=(\omega-n)/2$ \citep{Sedov46,vonNeumann47,Taylor50}. For $n\le\omega\le\omega_{g}(\gamma,n)$, and $n=2,3$, the self-similar exponent satisfies $\delta=0$ \citep{KushnirGap}, and for $\omega_{g}(\gamma,n)<\omega$ the self-similar exponent was found by \citet{Waxman93}. Panel (d): the shock velocity (normalized by $v_{0}$) as function of $\omega$ for $\gamma=5/3$ (solid line) and $\gamma=4/3$ (dashed line) for $n=2,3$. The inset zooms in around the ``gap'' region to make it clear that the shock velocity is finite there (and crosses smoothly the point $\omega=n$, shown as dotted line).}
   \label{fig:DeltaOmega}
\end{figure}

Generalizing the results of \citet{Waxman93,Waxman2010} to $n=1,2$ and of \citet{KushnirGap} to $n=2$ is straight forward, and the details are not given here. The results are that in the ``gap'' the shock propagates at a constant velocity ($\delta=0$) and $\xi_{c}(\Rsh)$ is a $C_{0}$ characteristic. Note that in this case the solution complies with the global conservation laws, since near the piston the flow deviates from the self-similar flow and the arguments given above for the divergence of $I_{k}$ do not hold. For $\omega>\omega_{g}(\gamma,n)$ the shock accelrates ($\delta>0$), $\xi_{c}(\Rsh)$ is a $C_{+}$ characteristic, and the solution passes thorough a singular point. The self-similar exponent $\delta$ for $\gamma=4/3$ and $\gamma=5/3$ as function of $\omega$ and $n$ is plotted in Figure~\ref{fig:DeltaOmega} for both the piston problem and the strong explosion problem.

The behavior of the solution in the ``gap'' region shows some interesting properties. Since $\delta=0$ and  $\xi_{c}(\Rsh)$ approaches the piston (see Equation~\eqref{eq:characteristics}), we can infer the shock velocity $\Vsh=v_{0}/\xi_{p}$ (there is an analog for $v_{0}$ for the strong explosion problem, since for $\omega>n$ most of the mass is near the explosion center and acquires a typical velocity). As can be seen in panel (d) of Figure~\ref{fig:DeltaOmega}, $\Vsh/v_{0}$ is some finite value (which depends on $\gamma$) at $\omega=\omega_{g}(\gamma,n)$. In other words, the transition to accelerating shocks happens from a finite ratio of $\Vsh/v_{0}$. Analyzing the solutions near the transition, we find that the transition happens as the causal connection between the piston and the shock is lost. This is natural, since for $\delta>0$ there cannot be such a causal connection (as explained in the beginning of the appendix), and indeed these solutions include a sonic point. The loss of the causal connection happens differently for different geometries. For $n=1,2$ the pressure at the piston $P(\xi_{p})$ approaches zero as  $\omega$ approaches $\omega_{g}(\gamma,n)$. For $n=3$ the pressure at the piston approaches some finite (nonzero) value at these limit, and the loss of connection is due to some $\xi_{a}(\gamma,n)>\xi_{p}$, for which $U(\xi_{a})+C(\xi_{a})=1$ (and therefore the time for a sound wave to cross this point is infinite).

% -------------------------------- Appendix B --------------------------
\section{B. Numerical 1D planar Lagrangian scheme}
\label{sec:1Ddetails}

The cells are separated into two groups based on their proximity to the contact surface. Cells $i=1,2,... i_{\rm pr}$ closer than a chosen Lagrangian point $m_{\rm pr}$ ,which is chosen as one of the nodes (node $i_{\rm pr}$ between cell $i_{\rm pr}$ and $i_{\rm pr}+1$), are assumed to have a uniform pressure $p_{\rm pr}$ and are treated separately from the rest of the cells. Other cells, $i>i_{\rm pr}$, are advanced by a standard scheme. During each time step, properties of the cells $i\leq i_{\rm pr}$ are calculated as follows: for a given (unknown) next step value of the pressure $p_{\rm pr}(t+\Delta t)$, the new densities $\rho_i(t+\Delta t)$ can be calculated for each cell using the equation of state. Using the constant cell masses, $m_i$, and new densities $\rho_i(t+\Delta t)$, the total length of the region can be calculated. By comparing this length to that obtained from the velocity and force equations on the border node $i_{\rm pr}$, an algebraic equation is obtained for $p_{\rm pr}(t+\Delta t)$, which is solved iteratively. We next write down this equation explicitly. Fore simplicity, we ignore nuclear burning which is straight forward to incorporate.

We assume that the thermodynamic variables $\rho_{i}(t)$ and $c_{s,i}(t)$ are defined at the cells' centers. The force equation for node $i_{\rm pr}$ can be solved similarly as to a regular node ($i>i_{\rm pr}$), and therefore $r_{i_{\rm pr}}(t+\Delta t)$ is known. The task is to solve for all other variables of the uniform pressure region at the time $t+\Delta t$. The length of the region (or equivalently the position of the node $i_{\rm pr}$) at $t+\Delta t$ is given by

\begin{equation}\label{eq:scheme1}
    z_{\rm pr}(t+\Delta t)=\sum_{i=1}^{i_{\rm pr}}\Delta z_{i}(t+\Delta t)=\sum_{i=1}^{i_{\rm pr}}\frac{m_{i}}{\rho_{i}(t+\Delta t)},
\end{equation}
and similarly
\begin{equation}\label{eq:scheme2}
    \sum_{i=1}^{i_{\rm pr}}\frac{m_{i}}{\rho_{i}(t)}=r_{i_{\rm pr}}(t),
\end{equation}
leading to
\begin{eqnarray}
\label{eq:scheme3}
    r_{i_{\rm pr}}(t+\Delta t)-r_{i_{\rm pr}}(t)=\sum_{i=1}^{i_{\rm pr}}m_{i}\left(\frac{1}{\rho_{i}(t+\Delta t)}-\frac{1}{\rho_{i}(t)}\right)
    \simeq-\sum_{i=1}^{i_{\rm pr}}\frac{m_{i}}{\left(\rho^{2}\right)_{i}(t+\Delta t/2)}\left(\rho_{i}(t+\Delta t)-\rho_{i}(t)\right),
\end{eqnarray}

In the uniform pressure region the flow is isentropic, and therefore
\begin{equation}\label{eq:scheme4}
    \rho_{i}(t+\Delta t)-\rho_{i}(t)=\frac{p(t+\Delta t)-p(t)}{\left(c_{s}^{2}\right)_{i}(t+\Delta t/2)}
\end{equation}
holds for every cell there. Using this with Equation~\eqref{eq:scheme3} we get
\begin{equation}\label{eq:scheme5}
    p(t+\Delta t)=p(t)-\frac{r_{i_{\rm pr}}(t+\Delta t)-r_{i_{\rm pr}}(t)}{\sum_{i=1}^{i_{\rm pr}}\frac{m_{i}}{\left(\rho^{2}c_{s}^{2}\right)_{i}(t+\Delta t/2)}}.
\end{equation}
Equations~\eqref{eq:scheme5} and~\eqref{eq:scheme4} are solved with iterations
\begin{eqnarray}\label{eq:scheme_iter1}
    p_{k}(t+\Delta t)&=&p(t)-\frac{r_{i_{\rm pr}}(t+\Delta t)-r_{i_{\rm pr}}(t)}{\sum_{i=1}^{i_{\rm pr}}\frac{m_{i}}{\left(\rho^{2}c_{s}^{2}\right)_{i,k-1}(t+\Delta t/2)}},\nonumber \\
    \rho_{i,k}(t+\Delta t)&=&\rho_{i}(t)+\frac{p_{k}(t+\Delta t)-p(t)}{\left(c_{s}^{2}\right)_{i,k-1}(t+\Delta t/2)},
\end{eqnarray}
supplemented by the equation of state
\begin{eqnarray}\label{eq:scheme_iter2}
    c_{s,i,k}(t+\Delta t)=c_{s}(p_{k}(t+\Delta t),\rho_{i,k}(t+\Delta t))
\end{eqnarray}
and by
\begin{eqnarray}\label{eq:scheme_iter3}
    \rho_{i,k}(t+\Delta t/2)&=&\frac{1}{2}\left(\rho_{i,k}(t+\Delta t)+\rho_{i}(t)\right),\nonumber \\
    \left(c_{s}^{2}\right)_{i,k}(t+\Delta t/2)&=&\frac{1}{2}\left[\left(c_{s}^{2}\right)_{i,k}(t+\Delta t)+\left(c_{s}^{2}\right)_{i}(t)\right],\nonumber \\
    \left(\rho^{2}c_{s}^{2}\right)_{i,k}(t+\Delta t/2)&=&\frac{1}{2}\left[\left(\rho^{2}c_{s}^{2}\right)_{i,k}(t+\Delta t)+\left(\rho^{2}c_{s}^{2}\right)_{i}(t)\right],
\end{eqnarray}
where $k$ is the number of the iteration and for the initial guess, $k=0$, we use
\begin{eqnarray}\label{eq:scheme_iter4}
    \rho_{i,0}(t+\Delta t/2)&=&\rho_{i}(t),\nonumber \\
    \left(c_{s}^{2}\right)_{i,0}(t+\Delta t/2)&=&\left(c_{s}^{2}\right)_{i}(t),\nonumber \\
    \left(\rho^{2}c_{s}^{2}\right)_{i,0}(t+\Delta t/2)&=&\left(\rho^{2}c_{s}^{2}\right)_{i}(t).
\end{eqnarray}
Typically, a few iterations are sufficient for convergence.

We implemented this scheme in the 1D, Lagrangian version of the VULCAN code \citep[for details, see][]{Livne1993IMT}.

% -------------------------------- Appendix C --------------------------

%-----------------------------------------------------------------------------
% --------------------------      BIBLIOGRAPGHY ---------------------------
%-----------------------------------------------------------------------------
\bibliographystyle{apj}

% ------------------------------ End of bibliography --------------------

\end{document}